\documentclass[graybox]{svmult}

\usepackage{mathptmx}       % selects Times Roman as basic font
\usepackage{helvet}         % selects Helvetica as sans-serif font
\usepackage{courier}        % selects Courier as typewriter font
\usepackage{type1cm}        % activate if the above 3 fonts are
\usepackage{makeidx}         % allows index generation
\usepackage{graphicx}        % standard LaTeX graphics tool
\usepackage{multicol}        % used for the two-column index
\usepackage[bottom]{footmisc}% places footnotes at page bottom

\usepackage{amssymb,amsmath}
\usepackage{graphicx}
\usepackage{color}
\usepackage{dcolumn}
\usepackage{pgfplots}
\usepackage{booktabs}
\usepackage{multirow}
\usepackage{dcolumn}
\usepackage{amsmath}
\usepackage{mathtools}
\usepackage{amsfonts}
\usepackage{amssymb}
\usepackage{epstopdf}
\usepackage{bm}
\usepackage{siunitx}
\usepackage{braket}
\usepackage{enumitem}
\usepackage{soul}
\usepackage{ulem}
\usepackage{color}
\usepackage{transparent}
\usepackage{pifont}
\usepackage[font={small}]{caption}
\newcommand\br{\begin{eqnarray}}
\newcommand\er{\end{eqnarray}}
\newcommand\be{\begin{equation}}
\newcommand\ee{\end{equation}}

\newcommand\bc{\begin{center}}
\newcommand\ec{\end{center}}

\begin{document}

\title*{Dynamical Tension Strings with Target Scale Symmetry producing DE, DM and why 4D?}
% Use \titlerunning{Braneworlds from dynamical tension} for an abbreviated version of
\author{Eduardo Guendelman}
\institute{Eduardo Guendelman \at Physics Department, Ben Gurion University of the Negev, Beer Sheva, Israel, \email{guendel@bgu.ac.il}}

\maketitle%

\abstract{ In the modified measure formulation string or branes the tension appear as an additional dynamical degree of freedom . Furthermore in the presence of an additional background scalar field that couples to the strings and locally changes the tension, the tension field really dynamical and the theory has an intrinsic target space scale symmetry. When many types of strings probing
the same region of space are considered this tension scalar is constrained by the requirement of quantum conformal invariance. For the case of two types of strings probing the same region of space with different dynamically generated tensions,
there are two different metrics, associated to the different strings. Each of these metrics have to satisfy vacuum Einstein´s equations and the consistency of these two
Einstein´s equations determine the tension scalar. The universal metric, common to both strings generically does not satisfy Einstein´s equation . We review a case where two string dependent metrics considered here are flat space in Minkowski
space and Minkowski space after a special conformal transformation and leads to a well defined brane world solution. We review how the model avoid swampland constraints making treatments for Dark energy and inflation more realistic and how
strings with a different tension appear as Dark Matter to us. Since the Dark strings and since the visible strings share the space time, including the same compactification, and since the compactification determines the particle content, we argue that the
dark strings lead to Dark copies of the standard model. An argument that many copies will lead to 4D effective spacetime as a consequence of a target space scale symmetry restoration.
}
\section{Introduction}
\label{sec:1}
String  Theories have been considered by many physicists for some time as the leading candidate for the theory everything,  including gravity, the explanation of all the known particles that we know and all of their known interactions (and probably more) \cite{stringtheory}. According to some, one unpleasant feature of string theory as usually formulated is that it has a dimension full parameter, in fact, its fundamental parameter , which is the tension of the string. This is when formulated the most familiar way.
The consideration of the string tension as a dynamical variable, using the modified measures formalism, which was  previously used for a certain class of modified gravity theories under the name of Two Measures Theories or Non Riemannian Measures Theories, see for example \cite{d,b, Hehl, GKatz, DE, MODDM, Cordero, Hidden}.  It is also interesting to mention that the modified measure approach has also been used to construct braneworld scenarios 
\cite{modified_measures_branes} and that Modified Measure Theories could be the effective Theories of Causal fermion systems \cite{MMandCFS}

 In the  modified measure approach to string theory, where  rather than to put the string tension by hand it appears dynamically.

This approach has been studied in various previous works  \cite{a,c,supermod, cnish, T1, T2, T3, cosmologyandwarped}. See also the treatment by Townsend and collaborators for dynamical string tension \cite{xx,xxx}, \cite{cosmologyandwarped}, which does not involve changing integration measures, so it cannot be used to achieve the goals presented in this paper however.

When many types of strings probing the same region of space are considered this tension scalar is constrained by the requirement of quantum conformal invariance. For the case of two types of strings probing the same region of space with different dynamically generated tensions, there are two different metrics, associated to the different strings.  Each of these metrics have to satisfy vacuum Einstein´s equations and the consistency of these two Einstein´s equations determine the tension scalar. The universal metric, common to both strings generically does not satisfy Einstein´s equation. The two string dependent metrics considered here are flat space in Minkowski space and Minkowski space after a special conformal transformation. The solutions are Braneworld \cite{Lidhtlikeandbraneworld} where for example the strings are confined between two expanding bubbles separated by a very small distance at large times or two planes moving at the speed of light. We present one of these braneworld solutions. We review how the model avoid swampland constraints making treatments for Dark energy and inflation more realistic and how
strings with a different tension appear as Dark Matter to us \cite{DMFROMSTRINGSWITHDIFFERENTTENSION} . Since the Dark strings and since the visible strings share the same space time, including the same compactification. or braneworld scenario, or a combination of both, and since the compactification determines the particle content, we argue that the
dark strings lead to Dark copies of the standard model. An argument that many copies will lead to 4D effective spacetime as a consequence of a target space scale symmetry restoration.
we also discuss why there is no minimal length or Limiting temperature and why there should be a dynamical
zero point energy density.

%Instead of simply listing headings of different levels we recommend to
%let every heading be followed by at least a short passage of text.
%Further on please use the \LaTeX\ automatism for all your
%cross-references and citations.

%And please note that the first line of text that follows a heading
%is not indented, whereas the first lines of all subsequent
%paragraphs are.

\section{The Modified Measure String Theory }

The standard world sheet string sigma-model action using a world sheet metric is \cite{pol1}, \cite{pol2}, \cite{pol3},

\begin{equation}\label{eq:1}
S_{sigma-model} = -T\int d^2 \sigma \frac12 \sqrt{-\gamma} \gamma^{ab} \partial_a X^{\mu} \partial_b X^{\nu} g_{\mu \nu}.
\end{equation}

Here $\gamma^{ab}$ is the intrinsic Riemannian metric on the 2-dimensional string world sheet and 
$\gamma = det(\gamma_{ab})$; $g_{\mu \nu}$ denotes the Riemannian metric on the embedding spacetime. $T$ is a string tension, a dimension full scale introduced into the theory by hand. \\

Now instead of using the measure $\sqrt{-\gamma}$ ,  on the 2-dimensional world-sheet, in the framework of this theory two additional world sheet scalar fields $\varphi^i (i=1,2)$ are considered. A new measure density is introduced:

\begin{equation}
\Phi(\varphi) = \frac12 \epsilon_{ij}\epsilon^{ab} \partial_a \varphi^i \partial_b \varphi^j.
\end{equation}

There are no limitations on employing any other measure of integration different than $\sqrt{-\gamma}$. The only restriction is that it must be a density under arbitrary diffeomorphisms (reparametrizations) on the underlying spacetime manifold. The modified-measure theory is an example of such a theory. \\

Then the modified bosonic string action is (as formulated first in \cite{a} and latter discussed and generalized also in \cite{c}

\begin{equation} \label{Action Mod Measure String1}
S = -\int d^2 \sigma \Phi(\varphi)(\frac12 \gamma^{ab} \partial_a X^{\mu} \partial_b X^{\nu} g_{\mu\nu} - \frac{\epsilon^{ab}}{2\sqrt{-\gamma}}F_{ab}(A)),
\end{equation}

where $F_{ab}$ is the field-strength  of an auxiliary Abelian gauge field $A_a$: $F_{ab} = \partial_a A_b - \partial_b A_a$. \\

It is important to notice that the action (\ref{Action Mod Measure String1}) is invariant under conformal transformations of the internal metric combined with an internal diffeomorphism in the space of the measure fields scalars, 

\begin{equation} \label{conformal}
\gamma_{ab} \rightarrow j \gamma_{ab}, 
\end{equation}

\begin{equation} \label{diffeo} 
\varphi^i \rightarrow \varphi^{'i}= \varphi^{'i}(\varphi^i)
\end{equation}
such that 
\begin{equation} \label{measure diffeo} 
\Phi \rightarrow \Phi^{'}= j \Phi
\end{equation}

Here $j$ is the jacobian of the diffeomorphism in the internal measure fields which can be an arbitrary function of the world sheet space time coordinates, so this can called indeed a local conformal symmetry.

To check that the new action is consistent with the sigma-model one, let us derive the equations of motion of the action (\ref{Action Mod Measure String1}). \\

The variation with respect to $\varphi^i$ leads to the following equations of motion:

\begin{equation} \label{eq variation of measure fields}
\epsilon^{ab} \partial_b \varphi^i \partial_a (\gamma^{cd} \partial_c X^{\mu} \partial_d X^{\nu} g_{\mu\nu} - \frac{\epsilon^{cd}}{\sqrt{-\gamma}}F_{cd}) = 0.
\end{equation}

since $det(\epsilon^{ab} \partial_b \varphi^i )= \Phi$, assuming a non degenerate case ($\Phi \neq 0$), we obtain, 

\begin{equation} \label{eq:a}
\gamma^{cd} \partial_c X^{\mu} \partial_d X^{\nu} g_{\mu\nu} - \frac{\epsilon^{cd}}{\sqrt{-\gamma}}F_{cd} = M = const.
\end{equation}

The equations of motion with respect to $\gamma^{ab}$ are

\begin{equation} \label{eq:8}
T_{ab} = \partial_a X^{\mu} \partial_b X^{\nu} g_{\mu\nu} - \frac12 \gamma_{ab} \frac{\epsilon^{cd}}{\sqrt{-\gamma}}F_{cd}=0.
\end{equation}

One can see that these equations are the same as in the sigma-model formulation . Taking the trace of (\ref{eq:8}) we get that $M = 0$. By solving $\frac{\epsilon^{cd}}{\sqrt{-\gamma}}F_{cd}$ from (\ref{eq:a}) (with $M = 0$) we obtain the standard string eqs. \\

The emergence of the string tension is obtained by varying the action with respect to $A_a$:

\begin{equation} \label{Tension Mod Measure String1}
\epsilon^{ab} \partial_b (\frac{\Phi(\varphi)}{\sqrt{-\gamma}}) = 0.
\end{equation}

Then by integrating and comparing it with the standard action it is seen that

\begin{equation}
\frac{\Phi(\varphi)}{\sqrt{-\gamma}} = T.
\end{equation}

That is how the string tension $T$ is derived as a world sheet constant of integration opposite to the standard equation (\ref{eq:1}) where the tension is put in ad hoc.
Let us stress that the modified measure string theory action 
does not have any \textsl{ad hoc} fundamental scale parameters. associated with it. This can be generalized to incorporate super symmetry, see for example \cite{c}, \cite{cnish}, \cite{supermod} , \cite{T1}.
For other mechanisms for dynamical string tension generation from added string world sheet fields, see for example \cite{xx} and \cite{xxx}. However the fact that this string tension generation is a world sheet effect 
and not a universal uniform string tension generation effect for all strings has not been sufficiently emphasized before.

Notice that Each String  in its own world sheet determines its own  tension. Therefore the  tension may not be universal for all strings.

\subsection*{Introducing world sheet currents that couple to the internal gauge fields and locally change the tension}

If to the action of the string  we add a coupling
to a world-sheet current $j ^{a}$,  i.e. a term
\begin{equation}
    S _{\mathrm{current}}
    =
    \int d ^{2} \sigma
        A _{a}
        j ^{a}
    ,
\label{eq:bracuract}
\end{equation}
 then the variation of the total action with respect to $A _{a }$
gives
\begin{equation}
    \epsilon ^{a b}
    \partial _{a }
    \left(
        \frac{\Phi}{\sqrt{- \gamma}}
    \right)
    =
    j ^{b}
    .
\label{eq:gauvarbracurmodtotact}
\end{equation}
We thus see indeed that, in this case, the dynamical character of the
brane is crucial here.
\subsection*{How a world sheet current can naturally be induced by a bulk scalar field, the Tension Field}

Suppose that we have an external scalar field $\phi (x ^{\mu})$
defined in the bulk. From this field we can define the induced
conserved world-sheet current
\begin{equation}
    j^{b}
    =
    e \partial _{\mu} \phi
    \frac{\partial X ^{\mu}}{\partial \sigma ^{a}}
    \epsilon ^{a b}
    \equiv
    e \partial _{a} \phi
    \epsilon ^{a b}
    ,
\label{eq:curfroscafie}
\end{equation}
where $e$ is some coupling constant. The interaction of this current with the world sheet gauge field  is also invariant under local gauge transformations in the world sheet of the gauge fields
 $A _{a} \rightarrow A _{a} + \partial_{a}\lambda $.

For this case,  (\ref{eq:gauvarbracurmodtotact}) can be integrated to obtain
\begin{equation}
  T =  \frac{\Phi}{\sqrt{- \gamma}}
    =
    e \phi + T _{0}
    ,
\label{eq:solgauvarbracurmodtotact2}
\end{equation}

The constant of integration $T _{0}$ could represent the initial value of the tension before the string enters the region where the tension field is present. Notice that the interaction is metric independent since the internal gauge field does not transform under the the conformal transformations. This interaction does not therefore spoil the world sheet conformal transformation invariance in the case the field $\phi$ does not transform under this transformation.  Therefore, one way to dynamically change the string tension tension dynamically is through the introduction this new background field, that changes the value of the the tension locally in the world sheet \cite{Ansoldi} . This background field was then called the  ¨tension field¨ 
and used for many different scenarios \cite{cosmologyandwarped}, \cite{Escaping}, the results of which have been summarized in \cite{summary}  and most recently for the construction of braneworld scenarios in dynamical string tension theories \cite{Life} and
\cite{Lidhtlikeandbraneworld}
If we have a scalar field coupled to a string or a brane in the way described in the sub section above, i.e. through the current induced by the scalar field in the extended object,  according to eq. 
(\ref{eq:solgauvarbracurmodtotact2})
, so we have two sources for the variability of the tension when going from one string to the other: one is the integration constant  $T _{0}$ , which we now generalize to  $T _{i}$ which varies from string to string (labelled by $i$) and the other the local value of the scalar field, which produces also variations of the  tension even within the string or brane world sheet. As we discussed in the previous section, we can incorporate the result of the tension as a function of scalar field $\phi$, given as $e\phi+T_i$, for a string with the constant of integration $T_i$ by defining the action that produces the correct 
equations of motion for such string, adding also other background fields, the anti symmetric  two index field $A_{\mu \nu}$ that couples to $\epsilon^{ab}\partial_a X^{\mu} \partial_b X^{\nu}$
and the dilaton field $\varphi $ .
\begin{equation}\label{variablestringtensioneffectiveacton}
S_{i} = -\int d^2 \sigma (e\phi+T_i)\frac12 \sqrt{-\gamma} \gamma^{ab} \partial_a X^{\mu} \partial_b X^{\nu} g_{\mu \nu}
\end{equation}
\begin{equation*}
 + \int d^2 \sigma A_{\mu \nu}\epsilon^{ab}\partial_a X^{\mu} \partial_b X^{\nu}
+\int d^2 \sigma \sqrt{-\gamma}\varphi R .
\end{equation*}
Notice that if we had just one string, or if all strings will have the same constant of integration $T_i = T_0$. We will take  cases where the dilaton field is a constant or zero, and the antisymmetric two index tensor field is pure gauge or zero, then the demand of conformal invariance for $D=26$ becomes the demand that all the metrics
\begin{equation}\label{tensiondependentmetrics}
g^i_{\mu \nu} =  (e\phi+T_i)g_{\mu \nu}
\end{equation}
will satisfy simultaneously the vacuum Einstein´s equations. 
The interesting case to consider is when there are many strings with different $T_i$, let us consider the simplest case of two strings, labeled $1$ and $2$ with  $T_1 \neq  T_2$ , then we will have two Einstein´s equations, for $g^1_{\mu \nu} =  (e\phi+T_1)g_{\mu \nu}$ and for  $g^2_{\mu \nu} =  (e\phi+T_2)g_{\mu \nu}$, 

\begin{equation}\label{Einstein1}
R_{\mu \nu} (g^1_{\alpha \beta}) = 0 
\end{equation}
and , at the same time,
\begin{equation}\label{Einstein1}
  R_{\mu \nu} (g^2_{\alpha \beta}) = 0
\end{equation}

These two simultaneous conditions above impose a constraint on the tension field
 $\phi$, because the metrics $g^1_{\alpha \beta}$ and $g^2_{\alpha \beta}$ are conformally related, but Einstein´s equations are not conformally invariant, so the condition that Einstein´s equations hold  for both  $g^1_{\alpha \beta}$ and $g^2_{\alpha \beta}$
is highly non trivial. Then for these situations, we have,

\begin{equation}\label{relationbetweentensions}
e\phi+T_1 = \Omega^2(e\phi+T_2)
\end{equation}
 which leads to a solution for $e\phi$
 
\begin{equation}\label{solutionforphi}
e\phi  = \frac{\Omega^2T_2 -T_1}{1 - \Omega^2} 
\end{equation}
which leads to the tensions of the different strings to be
\begin{equation}\label{stringtension1}
 e\phi+T_1 = \frac{\Omega^2(T_2 -T_1)}{1 - \Omega^2} 
\end{equation}
and
  \begin{equation}\label{stringtension2}
 e\phi+T_2 = \frac{(T_2 -T_1)}{1 - \Omega^2} 
\end{equation}

Both tensions can be taken as positive if $T_2 -T_1$ is positive and $\Omega^2$ is also positive and less than $1$.

\subsection{Flat space in Minkowski coordinates and flat space after a special conformal transformation }

The flat spacetime in Minkowski coordinates is,

 \begin{equation}\label{Minkowski}
 ds_1^2 = \eta_{\alpha \beta} dx^{\alpha} dx^{\beta}
\end{equation}

where $ \eta_{\alpha \beta}$ is the standard Minkowski metric, with 
$ \eta_{00}= 1$, $ \eta_{0i}= 0 $ and $ \eta_{ij}= - \delta_{ij}$.
This is of course a solution of the vacuum Einstein´s equations.

We now consider the conformally transformed metric

 \begin{equation}\label{Conformally transformed Minkowski}
 ds_2^2 = \Omega(x)^2  \eta_{\alpha \beta} dx^{\alpha} dx^{\beta}
\end{equation}
where conformal factor coincides with that obtained from the special conformal transformation
\begin{equation}\label{ special conformal transformation}
x\prime ^{\mu} =  \frac{(x ^{\mu} +a ^{\mu} x^2)}{(1 +2 a_{\nu}x^{\nu} +   a^2 x^2)}
 \end{equation}
for a certain D vector $a_{\nu}$.  which gives $\Omega^2 =\frac{1}{( 1 +2 a_{\mu}x^{\mu} +   a^2 x^2)^2} $
In summary, we have two solutions for the Einstein´s equations,
 $g^1_{\alpha \beta}=\eta_{\alpha \beta}$ and 
 
 \begin{equation}\label{ conformally transformed metric}
 g^2_{\alpha \beta}= \Omega^2\eta_{\alpha \beta} =\frac{1}{( 1 +2 a_{\mu}x^{\mu} +   a^2 x^2)^2} \eta_{\alpha \beta}
 \end{equation}
 
 We can then study the evolution of the tensions using 
 $\Omega^2 =\frac{1}{( 1 +2 a_{\mu}x^{\mu} +  a^2 x^2)^2}$.
 We will consider the cases where  $a^2 \neq 0 $.
  Notice that in the standard string theory two strings with different
 tensions cannot interact, cannot split into two strings with different tensions, etc. . By contrast, in the dynamical string tension theory, these interactions are only triggered when the string tensions are different. Since conventional string interactions cannot change the tension of the strings, since splitting or joining strings does not change the tension, these interaction do not play a role in the dynamics of the tensions of the strings, as opposed to the new interactions that arise at the multi string level considered here.
 
 Braneworlds from Flat space in Minkowski coordinates and Flat space after a special conformal transformation can appear if for example we consider two vacuum metrics,

The first being a flat spacetime in Minkowski coordinates is,

 \begin{equation}\label{Minkowski}
 ds_1^2 = \eta_{\alpha \beta} dx^{\alpha} dx^{\beta}
\end{equation}

where $ \eta_{\alpha \beta}$ is the standard Minkowski metric, with 
$ \eta_{00}= 1$, $ \eta_{0i}= 0 $ and $ \eta_{ij}= - \delta_{ij}$.
This is of course a solution of the vacuum Einstein´s equations.

and for the second we consider the conformally transformed metric, \cite{Life},  \cite{Lidhtlikeandbraneworld}, 

 \begin{equation}\label{Conformally transformed Minkowski}
 ds_2^2 = \Omega(x)^2  \eta_{\alpha \beta} dx^{\alpha} dx^{\beta}
\end{equation}
where conformal factor coincides with that obtained from the special conformal transformation
\begin{equation}\label{ special conformal transformation}
x\prime ^{\mu} =  \frac{(x ^{\mu} +a ^{\mu} x^2)}{(1 +2 a_{\nu}x^{\nu} +   a^2 x^2)}
 \end{equation}
for a certain D vector $a_{\nu}$.  which gives $\Omega^2 =\frac{1}{( 1 +2 a_{\mu}x^{\mu} +   a^2 x^2)^2} $
In summary, we have two solutions for the Einstein´s equations,
 $g^1_{\alpha \beta}=\eta_{\alpha \beta}$ and 
 
 \begin{equation}\label{ conformally transformed metric}
 g^2_{\alpha \beta}= \Omega^2\eta_{\alpha \beta} =\frac{1}{( 1 +2 a_{\mu}x^{\mu} +   a^2 x^2)^2} \eta_{\alpha \beta}
 \end{equation}
 
 We can then study the evolution of the tensions using 
 $\Omega^2 =\frac{1}{( 1 +2 a_{\mu}x^{\mu} +  a^2 x^2)^2}$.
 We will consider the cases where  $a^2 = 0 $, but $a^2 \neq 0 $  have also been considered for braneworld scenarios see for example our review in \cite{outoftheswampland}.

 The result in all cases is a situation where two hypersufaces where the string tension goes to infinity .
  The strings are then confined to be inside these surfaces, giving rise to a braneworld scenario. The very large string tensions are associated to a large Planck mass, which in turn implies the weakening of the swampland constraints \cite{outoftheswampland}.

\section{ Non singular braneworlds with periodic compactification} 
In previous research we have studied situations where there are two surfaces where the string tensions approach infinity, so we argue these situations represent the emergence of a braneworld scenario. Here, by introducing a periodic compactification of one dimension, instead of a segment compactification  we achieve still a very big growth of the string tension up to a maximum value, which is not infinite however. This will help us also with our discussion of strings with a different tension as dark matter.

For this purpose, let us take $\Omega^2 =\frac{1}{( 1 +2 a_{\mu}x^{\mu} +  a^2 x^2)^2}$, with $a^{\mu} = (A, A, 0,0,....)$
so that $a^2 = 0 $, 
 The string tensions of the strings one and two are given by
    \begin{equation}\label{stringtension1forBraneworld}
 e\phi+T_1 = \frac{(T_2-T_1)( 1 +2 a_{\mu}x^{\mu} +  a^2 x^2)^2}{( 1 +2 a_{\mu}x^{\mu} +  a^2 x^2)^2-1}=  \frac{(T_2-T_1)( 1 +2 a_{\mu}x^{\mu} +  a^2 x^2)^2}{(2 a_{\mu}x^{\mu} +  a^2 x^2)(2+2 a_{\mu}x^{\mu} +  a^2 x^2)}
\end{equation}
  \begin{equation}\label{stringtension2forBraneworld}
 e\phi+T_2 = \frac{(T_2-T_1)}{( 1 +2 a_{\mu}x^{\mu} +  a^2 x^2)^2-1}=  \frac{(T_2-T_1)}{(2 a_{\mu}x^{\mu} +  a^2 x^2)(2+2 a_{\mu}x^{\mu} +  a^2 x^2)}
\end{equation}
using that  $a^{\mu} = (A, A, 0,0,....)$, since $a^2=0$ we get that the tensions are
    \begin{equation}\label{stringtension1forBraneworld1}
 e\phi+T_1 =   \frac{(T_2-T_1)( 1 +2 a_{\mu}x^{\mu}   )^2}{(2 a_{\mu}x^{\mu} )(2+2 a_{\mu}x^{\mu} )}=  \frac{(T_2-T_1)( 1 - 2A (x-t))^2}{-4A(x-t) )(1+ A(x-t) )}
\end{equation}
 and in a similar fashion, we get,
  \begin{equation}\label{stringtension2forBraneworld2}
 e\phi+T_2 =   \frac{(T_2-T_1)}{-4A(x-t) )(1+ A(x-t) )}
\end{equation}

Defining the light like variable $\Delta = x-t$, we see that if $A< 0$ and for $\frac{(T_2-T_1)}{-A}> 0$, these tensions are positive and approach infinity at $\Delta =0 $ and  $\Delta =-1/A >0  $ .

We could avoid the infinte tension strings by starting the  $\Delta$ interval not at zero, but at a small positive value and then ending it at a value a bit smaller so the tensions recover  their initial value. This results in an interval which is a bit smaller, and can be made periodic, since at the start of the interval and at the end the tensions have the same values. 

To see the length of the new, periodic interval, we see that the tension of the second string depends only on 
$\Delta (1+A\Delta)$, and let us start with a positive value of 
$\Delta = \Delta_1 >0 $, and at a certain value $\Delta = \Delta_2 $,
the string tensions become of the same as those at 
$\Delta = \Delta_1 $, the relevant equation so that the second string acquires again its initial value is,
$$\Delta_1 (1+A\Delta_1) = \Delta_2 (1+A\Delta_2)$$
which lead us to an equation that allow us to solve for $\Delta_2$,
$$(\Delta_1 -\Delta_2)(1 + A (\Delta_1 +\Delta_2 )) =0 $$
so the solution for $\Delta_2 \neq  \Delta_1 $ is

$$\Delta_2 = - 1/A - \Delta_1 <  - 1/A  $$
since by assumption $\Delta_1  > 0 $ so that means no tension singularities appear in the new, now periodic interval.
Notice that since the tension of the second string and that of the first string differ by a constant, both string tensions share then the same periodicity. $\Delta_1 $ is now interpreted as the ultraviolet regularization that avoids the infinite growth of the strings at the borders of the now periodic interval.
\section{Can Dynamical Tension String Theory recover the Swampland?} 
The standard string theory is argued generates a space of acceptable
theories and a ¨swampland¨
 a space of theories that cannot be correct \cite{Vafa}.
The crucial point is that the string tension in Dynamical tension strings can become very large, indeed even infinity
since there is a simple relation between the string tension and Newton´s constant that is  the dimensionful gravitational constant parameter has been found in \cite{Paul} ,

$$\pi T  = (\kappa^2)^{-\frac{1}{d-2}}  $$

and $\kappa^2 = 8\pi G =\frac{1}{M^2_P}  $. 

We see that as the string tension goes to infinity  the Planck scale $M_P$ goes to infinity and the dimensionful gravitational constant parameter goes to zero in what we have argued is the target space scale invariant states. 
 In a general setting, there are a few statements made where the Planck scale appears \cite{Ooguri}:

 1. Distance conjecture: the statement into the requirement that trans-Planckian excursions can not be allowed for
any fields present in the cosmological evolution.
 
$$ \Delta \phi /M_P < O(1) $$
with $M_P$ being the reduced Planck mass

2,  Due to the difficulties of consistently constructing the meta-stable de-Sitter vacua at the heart
of cosmology it has been further proposed a requirement on possible field potentials of theories in the
Landscape  \cite{Vafa} and in particular \cite{Ooguri}, given by
either 
$$ M_P  \frac{dV/d\phi}{V} > O(1) $$
or 

$$  - M^2_P  \frac{d^2V/d\phi^2}{V} > O(1) $$

Of course the dynamical string tension theory provides a very meaningful way to weaken these constraints, since at the points in space time where the string tension goes to infinity, the Planck scale also goes to infinity and the swampland constraints above do not provide any constraint indeed.

\section{Disappearance of standard string interactions between strings with different tensions. }
  We have see a new type of interactions between strings, in fact between string with different tensions, mediated  by the tension field, but at the same time, the standard interactions of strings disappear. This is because these standard interactions have been formulated only for strings with the same tension, so these kind of interactions, disappear now, since they consist of splitting or joining, etc. of strings  which only make sense for strings with the same tension. 
  For example for the model of the non singular braneworlds with periodic compactification, the difference between the tensions is always a constant, so there will never be standard string interactions between these two species of strings.
 Even a simpler model that does not involve a tension field, would show similar effect,
 The tension field should be necessary to show some interaction between strings with different tensions, although not the standard string  interactions.  Also the cosmological emergence of different tension strings should be explained.
  One should point out however that strings with all types of tensions contribute to the structure of space time. The space time string tension metrics are related by a conformal transformation, so , in this way the effects of one string type affects the common metric and this back reacts on the other space time metric. So both strings species gravitate, 
\section{ Emergence of a new model for Dark Matter}
  We then look at what we know about our universe. There is indeed a big sector of our universe that does not share standard model interactions with us, the dark sector.  But in the context of our findings here, we see that Dark matter to us may consist of matter made out of strings with different tensions because of the decoupling of standard string interactions for strings with different tensions as proposed in \cite{DMFROMSTRINGSWITHDIFFERENTTENSION}. 
The decoupling of the dark sector, or strings with different
string tensions to ours, could have interesting and unexpected consequences, like for example a different reheating era for the conventional matter and for the dark sector. Such scenario was studied in  \cite{Freese} and the model of Dark Matter as strings with a different tension exactly fits this scenario.
As pointed out in \cite{Freese} , the 
primordial nucleosynthesis (BBN) provides strong evidence that the early Universe contained a hot plasma of photons and baryons with a temperature T bigger than  MeV. However, the earliest probes of dark matter originate from much later times around the epoch of structure formation. 
\section{Producing Dark Copies of the Standard Model}

It seems that a framework for implementing a dark matter scenario that mimics the standard model, but with different parameters could be based on \cite{DMFROMSTRINGSWITHDIFFERENTTENSION} 

Since all strings share the same space time, they should share the same compactification, etc. so the Dark Strings are likely to organize themselves in a similar way as the visible matter, leading probably to copies of the standard model with different parameters, since the string tension is not the same,

This should be somewhat related to the models of dark matter that mimic the standard model. as in references \cite{DARKMATTERMIMIKINGSTANDARDMODEL} , \cite{Randall} , or the Dark baryon production black holes models , which is based on a mirror dark standard model  \cite{mirror}.
\section{Unbroken Target Space scale invariance and four dimensional space time}
Notice that the string theory, has world sheet conformal invariance at the classical level, and this world sheet conformal invariace is requires to be extended to the quantum level.

At the classical level, the ordinary string theory does not have  target space scale invariance, which is very much related to the fact
that there is a definite scale in the theory, the string tension.

Indeed, in the ordinary string theory, a scale transformation of the background metric

$$ g_{\mu \nu}  \rightarrow   \omega g_{mu \nu}$$
where $\omega $ is a constant, 
is not a symmetry of the Polyakov action, but in the dynamical tension string theory, this transformation is a symmetry provided the world sheet gauge fiends and the  measure transforms as
$$ A_{a}  \rightarrow   \omega A_{a}$$
 $$\Phi(\varphi) \rightarrow   \omega ^{-1} \Phi(\varphi)  $$
 and the tension field transforms in a similar way,
 $$\phi \rightarrow   \omega ^{-1} \phi $$
We want to consider for the effective theory a modified  measure theory, because then the gravitational part of the action, linear in the curvature scalar, can be scale invariant (the Riemannian measure of integration does not allow this possibility). The appropriate integration measure in the space of $D$ , similar to what we did for strings is then, 
scalar fields  $\varphi_{a}, (a=1,2,...D)$, that is (we call the coordinates relevant to the effective theory A, B,  C,...) to differentiate them from the spacetime of the fundamental theory, for which we use Greek indices, we  also use  $G^{AB}$  for the metric in the effective theory. 
\begin{equation}
dV = 
d\varphi_{1}\wedge 
d\varphi_{2}\wedge\ldots\wedge 
d\varphi_{D}\equiv\frac{\Phi}{D!}d^{D}x
\label{dV}
\end{equation}
where
\begin{equation}
\Phi \equiv \varepsilon_{a_{1}a_{2}\ldots a_{D}}
\varepsilon^{A_{1}A_{2}\ldots A_{D}}
(\partial_{A_{1}}\varphi_{a_{1}})
(\partial_{A_{2}}\varphi_{a_{2}}) \ldots
(\partial_{A_{D}}\varphi_{a_{D}}).
\label{Fi}
\end{equation}
Accordingly, the total action in the D-dimensional 
space-time should be written in the form
\begin{equation}
S=\int\Phi Ld^{D}x
    \label{Act1}
\end{equation}
Our choice for the total Lagrangian 
density is
\begin{equation}
L=-\frac{1}{\kappa}R(\Gamma,G)+L_{strings}
\label{L1}
\end{equation}
and as mentioned before, the $R$ contribution  is multiplied by the measure and this can be target space scale invariant provided $\Phi \rightarrow \omega ^{-1} \Phi  $, as $G^{AB} \rightarrow  \omega G^{AB}$ and this is true  in either first or second order formalisms, since in either case  $R\rightarrow  \omega R $ we will see if this linear transformation in $\omega$ is also possible for the string matter part, 
which is represented by $L_{strings}$ is the matter Lagrangian density for string matter,
%$R(\Gamma,G)$
%the 
%scalar  curvature is  given, in the  first order formalism by
%\begin{equation}
%R(\Gamma,G)=G^{AB}R_{AB}(\Gamma)
%\label{R}
%\end{equation}

%R_{AB}(\Gamma)=R^{C}_{ABC}(\Gamma)
%\label{RAB}
%\end{equation}
\begin{equation}
L_{string}= -T\int d\sigma
d\tau\frac{\delta^{D}(x-X(\sigma,\tau))}{\sqrt{-G}}
\sqrt{Det(G_{AB}X^{A}_{,a}X^{B}_{,b})}
        \label{String}
\end{equation}
where $\int L_{string}\sqrt{-G}d^{D}x$ would be the action of a string
embedded in a $D$-dimensional space-time in the standard theory; $a,b$ label coordinates in the string world sheet and $T$ is the string tension. 
Notice that under a transformation 
$G^{AB} \rightarrow  \omega G^{AB}$,
$L_{string}\rightarrow \omega^{(D-2)/2}L_{string}$ , therefore
concluding that $L_{string}$ is a homogeneous function of $G^{AB}$
of degree one, that is the Target space scale invariance of the effective theory is satisfied only if $D=4$.

It would appear that we  have introduced a dimension full string tension $T$ and a dimension full Planck scale through $\frac{1}{\kappa}$, but this is not  so, since through the transformation 
$G^{AB} \rightarrow  \omega G^{AB}$, $T \rightarrow  \omega T$
and  $\frac{1}{\kappa} \rightarrow  \omega \frac{1}{\kappa} $,
so, only the ratio of the string tension to the Planck scale is meaningful. This ratio will be dependent on details of the compactification to four dimensions, etc. Full details are given in a longer publication \cite{4D}.

\section{Many Dark Copies of the Standard model brings us closer to Unbroken Target Space scale invariance and 4D }
If we have many copies through the dark strings with different tensions, we obtain a situation much closer than we would have if we have just one sting tension, from the point of view of achieving an unbroken target space scale invariance. So the more dark strings, the closer we will be to the target space symmetric state.
\section{No minimal length or Limiting temperature and dynamical zero point energy }
 In \cite{Escaping} we  have studied the possibility of avoiding the Hagedorn Temperature, since this temperature is inversely proportional to the string tension and the string tension can approach arbitrarily large values now.  In \cite{Andreev} Andreev has discussed the need to avoid the Hagedorn temperature in order to obtain a behavior more in accordance to that of QCD, since in the real world there is no phase transition but an analytic crossover. If strings are indeed relevant for QCD then one has to show that a stringy description is also valid for high T and he has shown that this can still be achieved but in the context of string models, but then these string models have to be multi tension string models.
The minimal length is similarly inversely proportional to the string tension, \cite{Veneziano} and must disappear in regions where the string tension is very large. 

Furthermore, a dynamical string tension will lead to an dynamical non commutative coordinates parameter, leading in turn to a dynamical zero point energy \cite{DZPE}.
\section{Clashing with famous symmetries and a correspondendaces }
The standard string theory has duality symmetries and the
ADS/CFT correspondance. These symmetries or correspondances clash with the Target scale invariance as an unbroken or almost unbroken symmetry,  since any symmetry or correspondance that makes reference to a scale like the string tension or the Planck scale, or a specific Anti de Sitter space which is defined for a specific negative cosmological constant manifestly violates Target scale invariance,  although they may reappear as symmetries of a low energy theory of the target space  scale invariance, which is the solution where all strings have  the same constant string tension, which spontaneouly breaks the  target space  scale invariance in a maximal way. From our point of view, that alternative is not interesting, because then we are back to all the problems the standard string theory has and we do not have something new. 
\section*{Acknowledgments}
I want to thank the organizers of LT-16 for their invitation to participate in this wonderful conference. I also want to thank COST ACTION COSMOVERSE CA21136
Addressing observational tensions in cosmology with systematics and fundamental physics and COST Action CA23130 - Bridging high and low energies in search of quantum gravity (BridgeQG) for finantial support and to Zeeya Merali for a FQXi report on aspects of this research \cite{ZEEYA}.


\begin{thebibliography}{9}
\bibitem{stringtheory}  ¨Superstrings¨,
John H. Schwarz, Vols 1 and 2, World Scientific, 1985; M. B. Green, J. H. Schwarz and E. Witten, Superstring Theory, Cambridge University
Press, 1987.
\bibitem{d}
E.I. Guendelman, A.B. Kaganovich, Phys.Rev.D55:5970-5980 (1997)
\bibitem{b}
Scale invariance, new inflation and decaying Lambda terms, 
E.I. Guendelman, Mod.Phys.Lett.A14, 1043-1052 (1999)
\bibitem{GKatz} Inflation and Transition to a Slowly Accelerating Phase from S.S.B. of Scale Invariance,  E.I. Guendelman, O. Katz,  Class.Quant.Grav. 20 (2003) 1715-1728 • e-Print: gr-qc/0211095 [gr-qc]
\bibitem{Hehl}
Frank Gronwald, Uwe Muench, Alfredo Macias, Friedrich W. Hehl, Phys.Rev.D 58 (1998) 084021 • e-Print: gr-qc/9712063 [gr-qc]
\bibitem{DE}
Eduardo Guendelman, Ramón Herrera, Pedro Labrana, Emil Nissimov, Svetlana Pacheva, Gen.Rel.Grav. 47 (2015) 2, 10 • e-Print: 1408.5344 [gr-qc]
\bibitem{MODDM}
Eduardo Guendelman, Douglas Singleton, Nattapong Yongram, JCAP 11 (2012) 044 • e-Print: 1205.1056 [gr-qc]
\bibitem{Cordero}
R. Cordero, O.G. Miranda, M. Serrano-Crivelli, JCAP 07 (2019) 027 • e-Print: 1905.07352 [gr-qc]
\bibitem{Hidden}
Eduardo Guendelman, Emil Nissimov, Svetlana Pacheva, Eur.Phys.J.C 75 (2015) 10, 472 • e-Print: 1508.02008 [gr-qc]
\bibitem{modified_measures_branes}
Conformally invariant gauge theory of three-branes in 6-D and the cosmological constant,
E.I. Guendelman, E. Spallucci, Phys.Rev.D 70 (2004) 026003 • e-Print: hep-th/0311102 [hep-th], Conformally invariant brane world and the cosmological constant, 
E.I. Guendelman, Phys.Lett.B 580 (2004) 87-92 • e-Print: gr-qc/0303048 [gr-qc]
\bibitem{MMandCFS} Modified Measures as an Effective Theory for Causal Fermion Systems
Felix Finster, Eduardo Guendelman, Claudio F. Paganini, 
e-Print: 2303.16566 [gr-qc]
\bibitem{signedGR} Signed General Coordinate Invariance, the Linde Universe Multiplication and Baby Universe Creation, 
  Eduardo Guendelman, e-Print: 2304.04056 [gr-qc].
\bibitem{signedstrings} Strings versus Anti Strings in the inversion invariant or proper volume formulation
E. Guendelman, 
e-Print: 2304.09946 [hep-th]
\bibitem{a} E.I. Guendelman, Class.Quant.Grav. 17, 3673-3680 (2000)
\bibitem{c}
E.I. Guendelman, A.B. Kaganovich, E.Nissimov, S. Pacheva, Phys.Rev.D66:046003 (2002)
\bibitem{supermod}
E.I. Guendelman, Phys.Rev.D 63 (2001) 046006 • e-Print: hep-th/0006079 [hep-th]
\bibitem{cnish}
Hitoshi Nishino, Subhash Rajpoot, Phys.Lett.B 736 (2014) 350-355
e-Print: 1411.3805 [hep-th].
\bibitem{T1}
T.O. Vulfs, E.I. Guendelman, Annals Phys. 398 (2018) 138-145 • e-Print: 1709.01326 [hep-th]
\bibitem{T2}
T.O. Vulfs, E.I. Guendelman, Int.J.Mod.Phys.A 34 (2019) 31, 1950204 • e-Print: 1802.06431 [hep-th]
\bibitem{T3}
T.O. Vulfs, Ben Gurion University Ph.D Thesis, (2021), arXiv:2103.08979. 
\bibitem{cosmologyandwarped}   E.I. Guendelman, Cosmology and Warped Space Times in Dynamical String Tension Theories, e-Print: 2104.08875 
\bibitem{Escaping} E.I. Guendelman, Escaping the Hagedorn Temperature in Cosmology and Warped Spaces with Dynamical Tension Strings, e-Print: 2105.02279 [hep-th].
\bibitem{Andreev}Oleg Andreev, 	JHEP 0903:098,2009, e.print: 0807.1017
\bibitem{summary} Implications of the spectrum of dynamically generated string tension theories, E.I. Guendelman, Int.J.Mod.Phys.D 30 (2021) 14, 2142028 • e-Print: 2110.09199 [hep-th]
\bibitem{Life} Life of the homogeneous and isotropic universe in dynamical string tension theories,  E.I. Guendelman,  Eur.Phys.J.C 82 (2022) 10, 857, https://link.springer.com/article/10.1140/epjc/s10052-022-10837-5 .
\bibitem{Lidhtlikeandbraneworld} Light like segment compactification and braneworlds with dynamical string tension, Eduardo Guendelman,  Eur.Phys.J.C 81 (2021) 10, 886. 2107.08005 [hep-th] , https://link.springer.com/article/10.1140/epjc/s10052-021-09646-z .
\bibitem{xx}
P.K. Townsend,  Phys.Lett.B 277 (1992) 285-288.
\bibitem{xxx}
E. Bergshoeff, L.A.J. London, P.K. Townsend,
Class.Quant.Grav. 9 (1992) 2545-2556, Class. Quantum Grav. 9 (1992) 2545-2556  • e-Print: hep-th/9206026 [hep-th]
\bibitem{pol1} Deser, S. and Zumino,Phys. Lett. B65 , 369, (1976)
\bibitem{pol2} Brink , L., Di Vechia, P and Howe, S. ,Phys. Lett. B65, 471 , (1976)
\bibitem{pol3} Polyakov, A. M,, ,Phys. Lett. B103 ,207, (1980).
\bibitem{Ansoldi}
S. Ansoldi, E. I. Guendelman, E. Spallucci, Mod.Phys.Lett.A 21 (2006) 2055-2065 • e-Print: hep-th/0510200 [hep-th]
\bibitem{HOLOMORPHIC} Holomorphic general coordinate invariant modified measure gravitational theory, 
Eduardo Guendelman,  Annals Phys. 458 (2023) 169466 • e-Print: 2308.09246 [gr-qc],
\bibitem{HOLOMORPHICregularization}Holomorphic gravity and its regularization of Locally Signed Coordinate Invariance
Eduardo Guendelman, Essay awarded honorable mention in  the 2024 Gravity Research Foundation Competition, published in Int.J.Mod.Phys.D 33 (2024) 15, 2441001
e-Print: 2402.00140 [gr-qc]
\bibitem{Polchinski} Joseph Polchinski, String Theory, vol. 1 , Cambridge University Press (1998) ; some papers on strings with background fields are C. G.
Callan, D. Friedan, E. J. Martinec and M. J. Perry, Nucl. Phys. B 262 (1985) 593;
T. Banks, D. Nemeschansky and A. Sen, Nucl. Phys. B 277 (1986) 67.
\bibitem{RS} A Large mass hierarchy from a small extra dimension , Lisa Randall, Raman Sundrum,  Phys.Rev.Lett. 83 (1999) 3370-3373 • e-Print: hep-ph/9905221 [hep-ph]
\bibitem{Paul} Dynamical String Tension in String Theory with Spacetime Weyl Invariance
Itzhak Bars, Paul Steinhardt, Neil Turok,  Fortsch.Phys. 62 (2014) 901-920
e-Print: 1407.0992 [hep-th]
DOI: 10.1002/prop.201400059
\bibitem{Fyn:e-m} R. P. Feynman, 
\emph{Quantum Electrodynamics} p. 68 (W.A. Benjamin, Inc, London 1961).
\bibitem{Schwinger1}J. Schwinger
\emph{Phys. Rev} $\mathbf{82}, 664$ 1951.
\bibitem{Julian Schwinger} Julian Schwinger,  Particles and Sources, Phys.Rev., 152 (1966) 1219-1226.
\bibitem{Paul} Itzhak Bars, Paul Steinhardt, Neil Turok, Dynamical String Tension in String Theory with Spacetime Weyl Invariance
  Fortsch.Phys. 62 (2014) 901-920
e-Print: 1407.0992 [hep-th]
DOI: 10.1002/prop.201400059
\bibitem{outoftheswampland} E. I. Guendelman, Dynamical string tension theories with target space scale invariance SSB and restoration
Eur. Phys. J. C (2025) 85: 276
Eur. Phys. J. C (2025) 85: 615
https://doi.org/10.1140/epjc/s10052-025-14296-6
Regular Article - Theoretical Physics Katherine 
\bibitem{DMFROMSTRINGSWITHDIFFERENTTENSION} Guendelman, E.I. Strings with a different tension as dark matter. Eur. Phys. J. C 85, 671 (2025). https://doi.org/10.1140/epjc/s10052-025-14408-2
\bibitem{Freese} Katty Freese,  Martin Wolfgang Winkler,  Dark matter and gravitational waves from a dark big bang, Phys.Rev.D 107 (2023) 8, 083522 • e-Print: 2302.11579 [astro-ph.CO]

\bibitem{Vafa} C. Vafa, The String Landscape and the Swampland, Oct., 2005.
10.48550/arXiv.hep-th/0509212.
\bibitem{Ooguri} H. Ooguri, E. Palti, G. Shiu and C. Vafa, Distance and de Sitter conjectures on the Swampland,
Physics Letters B 788 (2019) 180 [1810.05506].


\bibitem{4D} E. I. Guendelman, Dynamical string tension theories with target space scale invariance leading to 4D, Eur. Phys. J. C (2025) 85: 615
https://doi.org/10.1140/epjc/s10052-025-14296-6. 
A shorter version of this paper won the Gravity Research Foundation Essays on Gravitation
for
2025.
\bibitem{DARKMATTERMIMIKINGSTANDARDMODEL}
Sandip Roy, Xuejian Shen, Mariangela Lisanti, David Curtin, Norman Murray, Philip F. Hopkins,
Simulating Atomic Dark Matter in Milky Way Analogues, 	arXiv:2304.09878 [astro-ph.GA]
\bibitem{Randall} L. Randall, contribution to Black Holes and Cosmology, Reyjavik,  https://indico.mpp.mpg.de/event/10969/,
based on  L. Randall et. al.,   to be submitted.
 \bibitem{mirror}
Stefano Profumo, 
Dark baryon black holes,  Phys.Rev.D 111 (2025) 9, 095010, 
e-Print: 2502.16439 [hep-ph]
DOI: 10.1103/PhysRevD.111.095010 
 \bibitem{Veneziano} G. Veneziano A Stringy Nature Needs Just Two Constants
  1986 EPL 2 199
\bibitem{DZPE} E. Guendelman and D. Singleton, manuscript in preparation.
\bibitem{ZEEYA} Zeeya Merali, Escaping String Theory Swampland, EurekAlert!, https://www.miragenews.com/escaping-string-theory-swampland-1474428/
\end{thebibliography}
\end{document}